\documentclass[fleqn,usenatbib]{mnras}

\usepackage[T1]{fontenc}
\usepackage{ae,aecompl}

\usepackage{graphicx}	% Including figure files
\usepackage{amsmath}	% Advanced maths commands
\usepackage{amssymb}	% Extra maths symbols

\title[Linear Polarimetry of Trappist-1]{Time-resolved image polarimetry of Trappist-1 during planetary transits}

\author[P. A. Miles-P\'aez et al.]{
P. A. Miles-P\'aez$^{1,2}$\thanks{E-mail: ppaez@uwo.ca}, M. R. Zapatero Osorio$^{3}$,
E. Pall\'e$^{4,5}$, and S. A. Metchev$^{1,6}$
\\
$^{1}$Department of Physics \& Astronomy and Centre for Planetary Science and Exploration, The University of Western Ontario, London,\\ ON N6A 3K7, Canada\\  
$^{2}$Steward Observatory and Department of Astronomy, University of Arizona, 933 N. Cherry Avenue, Tucson, AZ 85721, USA\\
$^{3}$Centro de Astrobiolog\'ia (CSIC-INTA), Carretera de Ajalvir km 4, 28850 Torrej\'on de Ardoz, Madrid, Spain\\
$^{4}$Instituto de Astrof\'isica de Canarias, Calle V\'ia L\'actea s/n, 38205 La Laguna, Tenerife, Spain\\
$^{5}$Dpt. de Astrof\'isica, Univ. de La Laguna, Avda. Astrof\'isico Francisco S\'anchez s$/$n, 38206 La Laguna, Tenerife, Spain\\
$^{6}$Department of Physics \& Astronomy, Stony Brook University, Stony Brook, NY 11794--3800, USA\\
}

\begin{document}
\label{firstpage}
\pagerange{\pageref{firstpage}--\pageref{lastpage}}
\maketitle

% Abstract of the paper
\begin{abstract}
We obtained linear polarization photometry ($J$-band) and low-resolution spectroscopy ($ZJ$-bands) of Trappist-1, which is a planetary system formed by an M8-type low-mass star and seven temperate, Earth-sized planets. The photopolarimetric monitoring campaign covered 6.5 h of continuous observations including one full transit of planet Trappist-1d and partial transits of Trappist-1b and e. The spectrophotometric data and the photometric light curve obtained over epochs with no planetary transits indicate that the low-mass star has very low level of linear polarization compatible with a null value. However, the ``in transit" observations reveal an enhanced linear polarization signal with peak values of $p^* = 0.1\,\%$ with a confidence level of 3 $\sigma$, particularly for the full transit of Trappist-1d, thus confirming that the atmosphere of the M8-type star is very likely dusty. Additional observations probing different atmospheric states of Trappist-1 are needed to confirm our findings, as the polarimetric signals involved are low. If confirmed, polarization observations of transiting planetary systems with central ultra-cool dwarfs can become a powerful tool for the characterization of the atmospheres of the host dwarfs and the validation of transiting planet candidates that cannot be corroborated by any other method.
\end{abstract}

% Select between one and six entries from the list of approved keywords.
% Don't make up new ones.

%%%%%%%%%%%%%%%%%%%%%%%%%%%%%%%%%%%%%%%%%%%%%%%%%% 

%%%%%%%%%%%%%%%%% BODY OF PAPER %%%%%%%%%%%%%%%%%%

\section{Introduction}\label{intro}

The M8 star Trappist-1 is so far the only ultra-cool dwarf with transiting planets known \citep{2016Natur.533..221G,2017Natur.542..456G}. It is thus a unique benchmark that may be used to explore the feasibility of new techniques for seeking and characterizing close-in exoplanets around very low-mass stars and brown dwarfs, which are typically too faint for precise radial velocity measurements with current instrumentation.

Trappist-1 has an effective temperature $T_{\rm eff} = 2516\pm41$ K \citep{2018ApJ...853...30V}; this is low enough for naturally forming liquid and solid condensates in the upper photosphere  \citep{1996ApJ...472L..37F,1996A&A...305L...1T,1996A&A...308L..29T,1997ApJ...480L..39J,1997ARA&A..35..137A,2001ApJ...556..357A}. These condensates, sometimes referred to as ``dusty" particles that can be organized into ``clouds'', are expected to produce linear polarization at optical and near-infrared wavelengths via scattering processes \citep{2001ApJ...561L.123S,2011ApJ...741...59D}. Observations reveal that linear polarization is measurable in some late-M and L dwarfs, which show typical polarimetric degrees smaller than 1\%~ in the $I$- and $J$-bands 
\citep{2002A&A...396L..35M,2005ApJ...621..445Z,2009A&A...502..929G,2011ApJ...740....4Z,2013A&A...556A.125M,2015A&A...580L..12M,2017MNRAS.466.3184M}. Because the net polarization of a purely spherical atmosphere is zero, the linear polarimetry detections are usually explained by the asymmetries introduced by rotation-induced oblateness and/or patchy, non-uniform distribution of atmospheric condensates.   

Another likely asymmetry is the presence of a planet transiting the dwarf's dusty disk \citep{2016AJ....152...98S,2018ApJ...861...41S}. According to the theory, the net non-zero, time-dependent polarization is maximum at the inner contacts of the planetary ingress and egress phases. The polarimetric light curve during the transit shows characteristic profiles depending on the planet-to-parent object size ratio and the orbital distance. Models also predict that the linear polarization caused by planetary transits is higher in the $J$-band, and can be even higher than what is produced by rapid rotation-induced oblateness. Here, we report on our $J$-band polarimetric monitoring of Trappist-1 during the full transit of planet d and partial transits of planets b and e.

\section{Observations and data analysis}\label{obs}

\subsection{Linear polarimetry imaging \label{imaging}}

We collected polarimetry imaging data of Trappist-1 on the night of 2017 September 29, when three transits (planets b, d, and e) occurred. We employed the $J$-band filter centered at 1.2508 $\mu$m (bandpass of 0.1544\,$\mu$m) and the Long-slit Intermediate Resolution Infrared Spectrograph \citep[LIRIS;][]{2004SPIE.5492.1094M} attached to the Cassegrain focus of the 4.2-m William Herschel telescope (WHT) on the Roque de los Muchachos Observatory. LIRIS is equipped with a 1024$\times$1024 HAWAII detector and has a plate scale of 0$\farcs$25 pixel$^{-1}$ projected onto the sky. This yields a field of view of $4\farcm27\times4\farcm27$. In polarimetric mode, two Wollaston prisms split the light of the targets into four simultaneous images corresponding to vectors 0, 90, 135, and 45 deg, each of them with a reduced field of view of $4\arcmin\times1\arcmin$ as shown in Figure 
\ref{fig_liris} in the Appendix \ref{ap1}. We also observed our target using two different retarder plates (see \citealt{2017MNRAS.466.3184M}), which leads to a more precise determination of the polarization degree. 

As a result of our monitoring campaign, we obtained a total of 420 consecutive LIRIS frames on Trappist-1 (210 linear polarimetry measurements) over a continuous time period of $\approx$6.5 h. Individual integrations ranging from 20 through 70 s were applied depending on the air mass and seeing conditions. Because Trappist-1 is a bright M8 dwarf in the $J$-band ($J = 11.35\pm0.02$ mag) and in order to maximize the time cadence of the data, we performed stare observations (no nodding cycles). This observing strategy allowed us to maintain one reference star (2MASS\,J23063600--0502216, Figure\,\ref{fig_liris}) within the LIRIS polarimetric field of view, which was convenient to construct the differential (intensity) light curve of Trappist-1 (Section~\ref{lc}). Raw data were reduced using the Image Reduction and Analysis Facility software ({\sc iraf\footnote{{\sc iraf} is distributed by the National Optical Astronomy Observatories, which are operated by the Association of Universities for Research in Astronomy, Inc., under cooperative agreement with the National Science Foundation.}}). Each polarimetric vector was flat-fielded separately using appropriate data acquired through the polarimetric optics and at very high air masses during dusk to avoid the strong polarization of the Sun-illuminated sky.

To compute the Stokes paremeters $q$ and $u$, we followed \citet[and references therein]{2013A&A...556A.125M,2017MNRAS.466.3184M}. We used circular apertures around Trappist-1 of radii in the interval 0.5--6 times the full-width-at-half maximum (FWHM) of the reduced images with a step of 0.5$\times$FWHM. The telluric sky contribution was removed from the circular apertures using a ring with a inner radius of 6$\times$FWHM and a width of 1$\times$FWHM. We determined that apertures of 2--4$\times$FWHM in size were adequate for measuring $q$ and $u$. To increase the signal-to-noise ratio (S/N) of the polarimetric data, we obtained the averaged $q$ and $u$ values for every 7 independent, consecutive polarimetric measurements. Error bars of $q$ and $u$ were then computed as the standard deviation of 7 measures divided by~$\sqrt[]{7}$. We investigated the likely relationship of the observed Stokes parameters with the air mass, FWHM, the X and Y position of the target centroid in each image, { and the telescope rotator angle}, but we did not find any obvious correlation { (see Appendix \ref{ap0})}.  The degree of linear polarization, $P$, and its associated uncertainty, $\sigma_P$, were derived from the quadratic sum of the Stokes parameters and their errors, respectively. For low values of $P$, and since $P$ is always a positive quantity, the polarimetric degree tends to be overestimated. We corrected for this effect using $p^{*}\,=\,\sqrt{P^2-\sigma^2_p}$ \citep{1974ApJ...194..249W,1985A&A...142..100S}, thus obtaining the debiased linear polarization degree $p^*$ ($p^* = 0.0^{+\sigma_p}_{-0}$ when $\sigma_p > P$). Table~\ref{table1} provides $q$, $u$, and $p^*$ along with their corresponding central time and observing time interval. We did not compute the vibration angle of polarization because the observed Stokes parameters are small and will lead to large angular uncertainties. Instead, we shall describe the results of the polarimetric monitoring of Trappist-1 in terms of $q$, $u$, and $p^*$.

On the same observing night as Trappist-1, we also acquired $J$-band linear polarimetry images of the unpolarized standard star HD\,14069 \citep{1990AJ.....99.1243T} for controlling the LIRIS instrumental polarization. Integrations of 4 s were employed per retarder plate. We obtained the following values: $q = 0.02\pm0.04\%$, $u = -0.03\pm0.04\%$, and $p^* = 0.00^{+0.06}_{-0.0}\%$, thus confirming that LIRIS has a very low instrumental polarization~at $J$-band wavelengths, with an upper limit of 0.18\%~(3 $\sigma$ confidence level).

\subsection{Linear spectropolarimetry \label{spectropol}}
We carried out low-resolution spectropolarimetry observations of Trappist-1 using LIRIS at the WHT on 2017 August 4 between $\approx$4 UT and $\approx$5 UT, in which no planetary transits were expected to take place. This is about 7 weeks prior to the monitoring of the planetary transits reported in section \ref{imaging}. In its spectropolarimetric mode, LIRIS is equipped with a narrow long slit that has 0\farcs75 width and $1'$ length. We employed the low resolution $ZJ$ grism. This instrumental configuration yields a wavelength coverage of 0.9--1.5 $\mu$m, a spectral resolution of 18.1 \AA~($R \approx 600$ at 1.1 $\mu$m), and a nominal dispersion of 6.06 \AA\,pixel$^{-1}$. For calibration purposes, the polarized star Cyg OB2 \#12 \citep{1992ApJ...386..562W} and the unpolarized star HD\,212311 \citep{1990AJ.....99.1243T} were also observed with the same instrumental configuration on 2017 August 3. The seeing was stable at around 0\farcs5 on the first night and variable between 0\farcs5 and 1\farcs2 on the second night. Sky transparency was excellent on both nights.

LIRIS spectra were acquired with the two retarder plates and at two nodding positions along the slit separated by 16\arcsec~for a proper subtraction of the Earth's sky contribution. Individual integration times were 900 s (Trappist-1), 180 s (HD\,212311) and 3 s (Cyg OB2 \#12). The complete polarimetric cycle of Trappist-1 lasted nearly 1 h. Both science and calibration targets were acquired on the same detector spot to minimize systematic errors. Data were not obtained at parallactic angle. Objects were acquired on the slit using the $J$-band filter, which is fully covered by the spectroscopic observations, i.e., we ensured that most of the flux passed through the slit. In addition, the LIRIS $Z$-band is centered at very red wavelengths, and we do not expect significant flux losses due to the Earth's atmospheric refraction.

Raw spectra were reduced with standard procedures within {\sc iraf}. The two nodding frames were subtracted, flat-fielded, and the four spectra per frame were optimally extracted. A full wavelength solution from calibration Argon lamps taken during the observing nights was applied separately to each of the four spectra. The error associated with the fifth-order Legendre polynomial fit to the wavelength calibration is typically 10--15\%~the nominal dispersion.  

We applied the flux ratio method \citep{2005ApJ...621..445Z} to derive the final spectropolarimetric data corresponding to the Stokes parameters $q$ and $u$ and the linear plarization degree $P$ shown in Figure~\ref{polspectra}. The bottom panel illustrates the polarized standard star Cyg OB2 \#12, whose linear polarization degree shows a marked dependency on wavelength. On average, we measured linear polarization degrees of $P$ = 3.83\,$\pm$\,0.01\%~ ($\sigma = 0.39\%$) and $P$ = 2.31\,$\pm$\,0.02\%~($\sigma = 0.44\%$) for Cyg OB2 \#12 after integrating our data over the LIRIS $Z$- and $J$-bands, respectively. These values are systematically $\approx$1\%~below those published by \citet{1992ApJ...386..562W}. A similar result is obtained by \citet[see their Figure~1]{steele17}, who determined a linear polarization degree at optical wavelengths (where the polarization of Cyg OB2 \#12 is maximum) smaller than the canonical one. Therefore, we assume that between \citet{1992ApJ...386..562W} and \citet{steele17} observations there has been a change in the polarizing conditions of the material surrounding this extinguished star, and we do not apply any additional correction to our LIRIS data. The observations of the unpolarized star revealed that there is no significant instrumental polarization with an upper limit at $\approx$0.2\%~over the entire wavelength coverage. This result is similar to that obtained from the polarimetry imaging (Section~\ref{imaging}).

The top panel of Figure~\ref{polspectra} shows the LIRIS spectrum of Trappist-1 with no telluric or instrumental response corrections. It is shown for a convenient illustration of the location of strong telluric features and intrinsic features of the M8V dwarf. The spectropolarimetric $q$ and $u$ parameters of Trappist-1 are depicted in the middle panel of Figure~\ref{polspectra}. The spectra are flat and reveal that Trappist-1 is not linearly polarized between 0.9 and 1.34 $\mu$m. The average values of the Stokes parameters are determined at $q = 0.00 \pm 0.45\%$ and $u = 0.00 \pm 0.50\%$, where the error bars account for the standard deviation of the corresponding spectra (the errors of the mean are $\approx$30 times smaller). These observations confirm that Trappist-1 is essentially unpolarized in the $J$-band wavelengths when none of its planets is transiting. This is an important result for the interpretation of the polarimetric light curve discussed in Section \ref{pol}.

\begin{figure}
\centering
\includegraphics[width=0.45\textwidth]{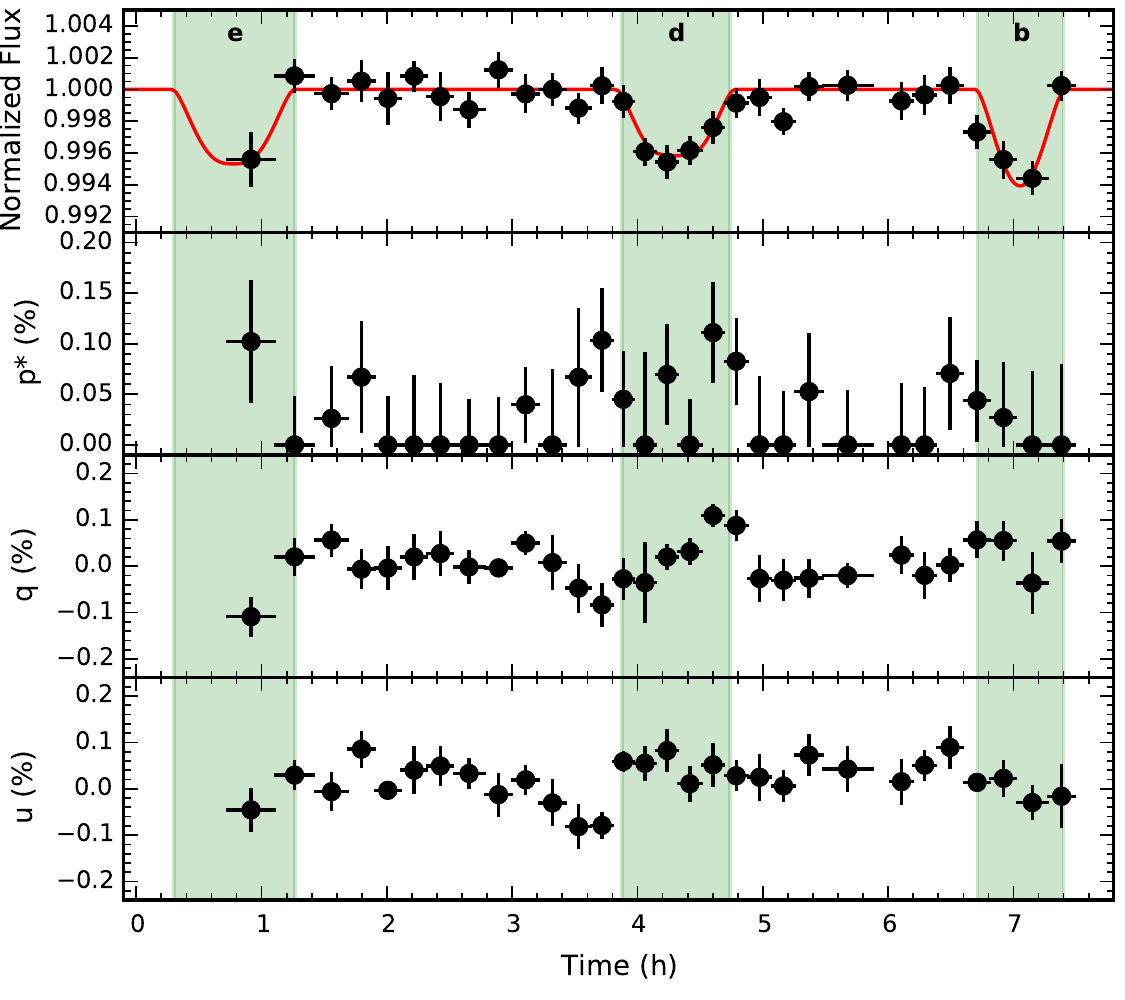}
\caption{From top to bottom: Trappist-1 $J$-band intensity (differential), linear polarization, Stokes $q$ and $u$ parameters light curves taken on 2017 Sep 29. Vertical error bars stand for the uncertainties of the average measurements (see text) and horizontal error bars indicate the time coverage accounted by the observations. The red line (top panel) shows the best fit of the intensity light curve including the transits of planets Trappist-1e, d, and b derived with EXOFAST. Greenish strips denote the duration of the planetary transits according to the EXOFAST solution for the intensity light curve. We adopted the Modified Julian Date 58025.81434 as the zero time in all the panels.}
\label{fig1}
\end{figure}

\subsection{Intensity light curve}\label{lc}
We derived the LIRIS intensity light curve of Trappist-1 (data described in Section \ref{imaging}) by means of differential photometry. For each of the four polarimetric vectors in a LIRIS frame, we computed the flux of Trappist-1 using a circular aperture of radius  2.5$\times$FWHM and a sky ring with an inner annulus and width of 3.5$\times$ and 1$\times$FWHM, respectively. Then we added the fluxes of the vectors to obtain the total flux of Trappist-1 per frame. We repeated the same process for the reference star 2MASS J23063600--0502216, which is the only relatively bright source in the LIRIS field of view besides our target (Figure \ref{fig_liris}). Trappist-1's intensity (or differential) light curve was obtained by deriving the flux ratio between the target and the reference star. We measured the average value of the differential light curve at epochs with no planetary transits for detrending and normalizing the curve to unity. In order to have the same temporal resolution as the polarization light curve, we averaged the differential photometry every 14 consecutive, independent measurements. Photometric uncertainties were then estimated as the errors of the mean photometry. The final, normalized intensity light curve is illustrated in the top panel of Figure~\ref{fig1}; measurements are provided in the last column of Table \ref{table1}. The data covered the full transit of Trappist-1d and partially those of Trappist-1b and e based on the ephemeris of \citet{2017Natur.542..456G}. 

We fit the duration, depth, and central time of the transits of planets b and d, and only the depth and central time for planet e by using EXOFAST \citep{2013PASP..125...83E}; the remaining parameters involved in the light curve model such as orbital inclination angle, stellar radius, and limb-darkening coefficients, were fixed using those listed in \citet{2017Natur.542..456G}. The measured values are given in Table \ref{table1} and the best fit solution is plotted in the top panel of Figure~\ref{fig1}. The mid-transit times derived from our best-fit solution were converted from Universal Time to Barycentric Julian Date (BJD$_{\rm TDB}$) using the calculator provided by \citet{2010PASP..122..935E}. We found time differences between the predicted transit dates from the ephemeris \citep{2017Natur.542..456G} and our derivations of $-12.2\pm3.0$, $10.5\pm1.6$, and $3.3\pm8.0$ min for Trappist-1b, d, and e, respectively. The transit time differences of planets d and e lie within 2.5 $\sigma$ with respect to the ephemeris and can be explained by the quoted error bars and the computed corresponding transit time variations (TTVs, see extended data Figure 4 of \citealt{2017Natur.542..456G}). The situation is different for planet b, where the difference of $-12.2\pm3.0$ min is incredibly high and cannot be accounted by quoted uncertainties and expected TTVs. On the contrary, we determined transit depths and durations for all three planets that coincide with those of \citet{2017Natur.542..456G} at the 1 $\sigma$ level, although our determinations tend to be systematically deeper (transit depth) and longer (transit duration) probably because of the poor time resolution of the LIRIS $J$-band data.

\begin{figure}
\centering
\includegraphics[width=0.35\textwidth]{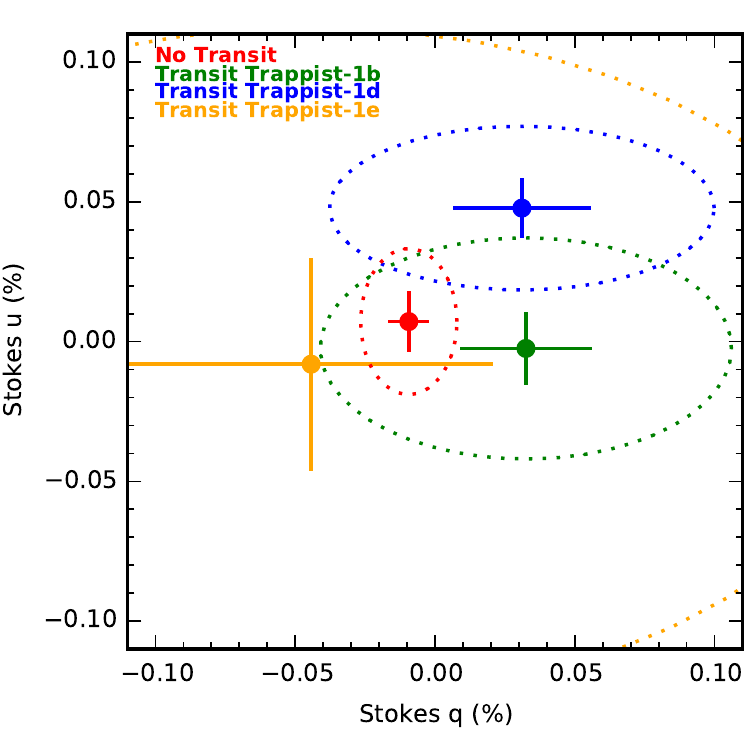}
\caption{Average Stokes parameters $q$ and $u$ during the transits of planets Trappist-1d (blue), b (green), and e (orange). The red dot stands for the average Stokes parameters of the central star when no planet is transiting. The dotted ellipses show the 3 $\sigma$ confidence interval using the projection method described in \citet{1986VA.....29...27C}. {Vertical and horizontal error bars stand for 1 $\sigma$ uncertainties.}
}
\label{fig2}
\end{figure}

\section{Linear polarization light curve} \label{pol}

The three bottom panels of Figure~\ref{fig1} illustrate the light curves corresponding to the $J$-band linear polarimetry photometry: $p^*$ and the Stokes parameters $q$ and $u$. At epochs with no planetary transits, the light curves show a flat behavior near zero polarization with a small dispersion, indicating that the disk-integrated polarization of Trappist-1 star remains below the $J$-band detectability limit of LIRIS. This fully agrees with the spectropolarimetric observations presented in Section~\ref{spectropol}. The polarimetric light curves typically show higher values of $p^*$ and a sensibly larger dispersion of the polarimetric data during the three planetary transits (marked with a greenish area in Figure~\ref{fig1}). 

To study the statistical significance of the polarimetric detections and non-detections in and out of planetary transits in the Trappist-1 system, we computed the intervals of confidence for the Stokes parameters $q$ and $u$ using the (rather conservative) ``projection method'' \citep[equations 18--20 of][]{1986VA.....29...27C}, although other procedures to determine the confidence intervals of polarimetric measurements are available in the literature \citep{1974apoi.book..361S,1985A&A...142..100S}. This study focused on the Stokes parameters $q$ and $u$ because they have associated errors that follow a Gaussian distribution, thus making the derivation of confidence intervals easier, as opposed to the errors of $p^*$, which follow a Rice distribution when very low polarimetric signals are involved. For each planet Trappist-1b, d, and e, the average $q$ and $u$ values were determined during the transits (after the transit starting and end times given by the model plotted on the top panel of Figure~\ref{fig1}). For the epochs with no planetary transits, we also derived the average Stokes parameters finding $q=0.009\pm0.006\,\%$ and $u=0.007\pm0.010\,\%$, which are consistent with a null linear polarization at the $\approx$1 $\sigma$ level. All these mean values (in and out of transit) were given their 3 $\sigma$ confidence intervals following \citet{1986VA.....29...27C}; they are listed in Table \ref{table1} and shown in the $q-u$ plane of Figure~\ref{fig2}. 

For the number of data points covering the full transit, planet Trappist-1d offers the best opportunity to confirm the presence of detectable linear polarization in the atmosphere of its parent M8-type star. During Trappist-1d transit, the Stokes parameter values were determined at $q=0.031\pm0.024\,\%$ and $u=0.048\pm0.010\,\%$ (shown with a blue dot in Figure~\ref{fig2}). They clearly deviate from the null polarization at the 3 $\sigma$ confidence level. This result provides a solid confirmation of the linearly polarized atmosphere of Trappist-1 and supports the theory of \citet{2016AJ....152...98S}. The transits of Trappist-1e and b were not optimally observed for this project: partial transit coverages and low number of data points acquired during transit, which strongly affects the 3 $\sigma$ confidence levels of the in-transit mean polarimetric values. Nevertheless, the observations of planet b do not contradict the conclusions reached with Trappist-1d in Figure~\ref{fig2}.

Very noticeable is, however, the different transit duration between the polarimetric and the intensity light curves (Figure~\ref{fig1}): the deviation of $p^*$, $q$, and $u$ from the null value seems to start 9$\pm$4 min before Trappist-1d's ingress and goes 7$\pm$4 min beyond the planet's egress as determined from the best-fit model of the intensity/differential light curve; this is, Trappist-1d transit, as seen from the polarimetric light curve, appears to have longer duration by about $16 \pm 8$ min. This represents 33$\pm$16\%~more prolonged transit than that tabulated by \citet{2016Natur.533..221G,2017Natur.542..456G}. We explored whether this feature could be an artifact produced by the binning of the data and found that it is present independently of the binning size and temporal resolution. We remark that the same effect is seen after and before the transits of Trappist-1e and b, respectively. Also, there is one non-zero value of p* (with a low significance of $p^*/\sigma_P=1.4$) close to the central part of the transit of Trappist-1d, which breaks the flatness predicted by the theory. These features do not invalidate our conclusions about the net change in the polarimetric signal of Trappist-1 during its transits based on the Stokes parameters (Figure \ref{fig2}), which are  directly measurable quantities.

\section{Discussion and final remarks}

The net change of polarimetry observed during the transit of Trappist-1d (Fig. \ref{fig2}) qualitatively coincides with the theoretical predictions of \citet[][see his Figure~1]{2016AJ....152...98S}, thus providing support to the theory. When the planet is passing in front of the stellar disk, there is a change in the apparent disk-integrated polarization of the star, which reaches its maximum at the inner ingress and egress contact points (when the geometrical asymmetry is the largest) and a minimum at mid-transit time (when the polarization of the borders of the star is self-cancelled). Quantitatively, Sengupta's models can predict peaks of linear polarization degrees of about 0.1\,\%~in the $J$-band for the inner planets of Trappist-1 system { \citep{2018ApJ...861...41S}} since the central star is classified as an M8. This 
fully agrees with our measurements. Although our data are not accurate enough to properly delineate the predicted peaks at ingress/egress and the flat signal of polarization in the central part of the transit. Much higher polarization degrees (e.g., $\approx$1\,\%) are expected for early-L dwarfs because in principle these dwarfs have dustier atmospheres. The major difference between the observations reported here and the theory lies in the time predictions of the critical moments for maximum polarization.

The detected linear polarization is likely originated in Trappist-1's stellar atmosphere or in the immediate surroundings of the star. The most likely physical scenario causing this polarization is the presence of dusty particles in the upper photosphere resulting from the natural condensation processes of volatile elements at low temperatures and high pressures (see references in Section \ref{intro}). That the polarization degree is close to zero when there are no planetary transits can be explained by an homogeneous distribution of the dust over the stellar disk. It may also imply that the amounts of atmospheric dust are not strongly affected/modified by the intense X-ray and ultraviolet radiation of the star \citep{2017MNRAS.465L..74W} and stellar activity. In addition, Trappist-1 is a slow rotator with a rotation period determined at $3.30\pm0.14$ d \citep{2017NatAs...1E.129L}, although it has been recently suggested that this period may not be a rotational signal but rather a characteristic timescale of active regions \citep{2018ApJ...857...39M}. Therefore, the oblateness index should be rather small, which favors the spherical symmetry of Trappist-1's atmosphere and the self-cancelation of the net linear polarization over the visible disk. 

Any scenario that assumes that the polarizing region is contained within one stellar radius, however, cannot account for the observed transit duration between the intensity and polarimetric light curves. The polarizing dust would have to be located at distances of 1.33$\pm$0.16 stellar radius from the center of the star to explain the long duration of the observed ``polarimetric transits". This would imply an extended, faint dusty atmosphere which does not contribute substantially to the disk-integrated intensity of the star, but with a high impact in the polarization curve. Nevertheless, given the fact that we only have a single full transit and two partial transits, this result is rather speculative. Additional polarimetric data monitoring for various transits of different planets are needed to confirm with a higher level of confidence the results reported in this work, particularly the one related to the longer duration of the transits.

Our observations suggest that linear polarization may become a powerful tool to validate transiting planets around ultra-cool dwarfs, especially the dustier and fainter L dwarfs. In addition, multi-band linear polarization observations can be used to infer the typical size of the dusty particles, and to investigate the activity cycles of the polarized central dwarf and their impact on the habitability conditions of the surrounding planets. 

\section*{Acknowledgements}

This article is based on observations made in the Observatorios de Canarias del IAC with the William Herschel Telescope (WHT) operated on the island of La Palma by the Isaac Newton Group in the Observatorio del Roque de los Muchachos. This work is partly financed by the Spanish Ministry of Economics and Competitiveness through grants ESP2013-48391-C4-2-R and AYA2016-79425-C3-2-P.

%%%%%%%%%%%%%%%%%%%%%%%%%%%%%%%%%%%%%%%%%%%%%%%%%%

%%%%%%%%%%%%%%%%%%%% REFERENCES %%%%%%%%%%%%%%%%%%

% The best way to enter references is to use BibTeX:

\bibliographystyle{mnras}
\bibliography{biblio} % if your bibtex file is called example.bib

% Alternatively you could enter them by hand, like this:
% This method is tedious and prone to error if you have lots of references
%\begin{thebibliography}{99}
%\bibitem[\protect\citeauthoryear{Author}{2012}]{Author2012}
%Author A.~N., 2013, Journal of Improbable Astronomy, 1, 1
%\bibitem[\protect\citeauthoryear{Others}{2013}]{Others2013}
%Others S., 2012, Journal of Interesting Stuff, 17, 198
%\end{thebibliography}

%%%%%%%%%%%%%%%%%%%%%%%%%%%%%%%%%%%%%%%%%%%%%%%%%%

%%%%%%%%%%%%%%%%% APPENDICES %%%%%%%%%%%%%%%%%%%%%

% Don't change these lines
\bsp	% typesetting comment
\label{lastpage}

\appendix

\section{Additional figures and observed values for Trappist-1.}\label{ap1}

\begin{figure}
\centering
\includegraphics[width=0.48\textwidth]{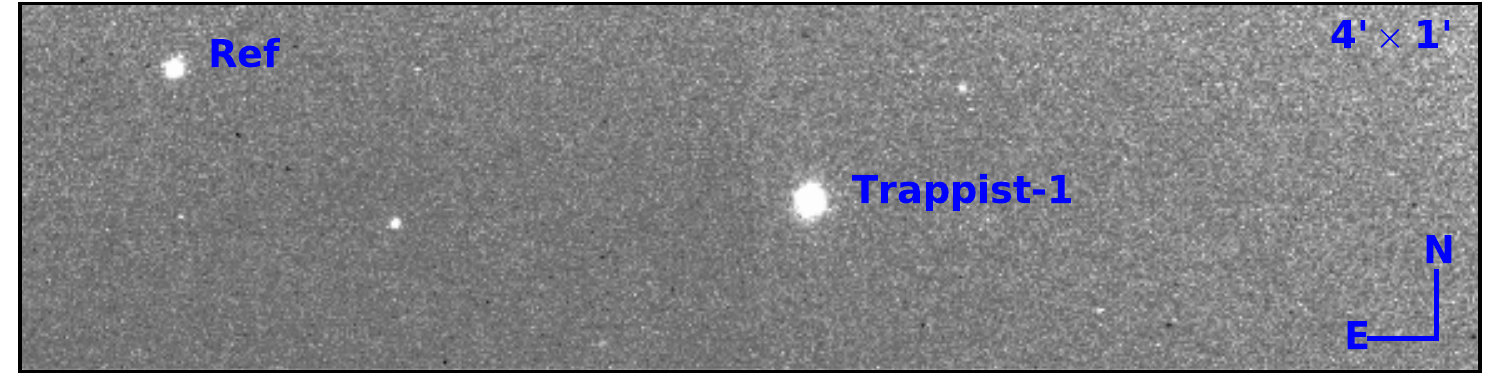}
\caption{LIRIS field of view in polarimetric mode. This slice corresponds to one of the four polarimetric images registered in a LIRIS frame. Our target (Trappist-1) and a reference star (2MASS\,J23063600--0502216) are labeled. Orientation and field size are indicated.}
\label{fig_liris}
\end{figure}

\begin{figure}
\centering
\includegraphics[width=0.48\textwidth]{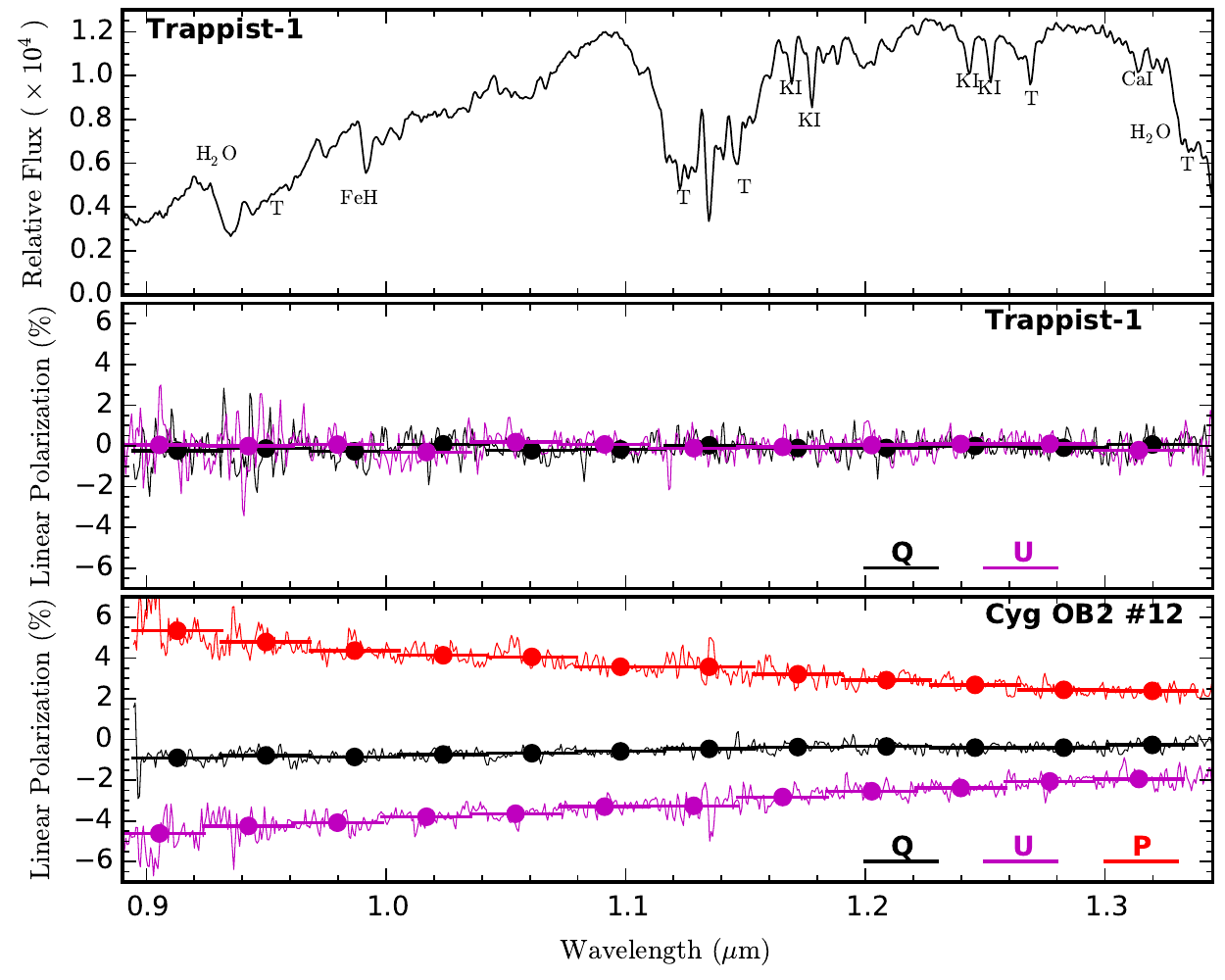}
\caption{{\sl Top:} LIRIS spectrum of Trappist-1. No telluric or instrumental response corrections have been applied. Some molecular and atomic features are identified (``T" stands for significant telluric contribution). {\sl Middle:} spectropolarimetric data of Trappist-1. The Stokes $q$ and $u$ parameters are depicted with black and magenta lines, respectively. The solid dots correspond to the binned spectra with binsize of 36 pixels. {\sl Bottom:} spectropolarimetric observations of the polarized standard star Cyg OB2 \#12. The red, black and magenta lines and solid dots indicate the linear polarization degree $P$, and Stokes $q$ and $u$ parameters, respectively. The horizontal error bars represent the bin size of the binned data (solid dots). }
\label{polspectra}
\end{figure}

%%%Table
\begin{table*}
\caption{Observed values of intensity and linear polarization for Trappist-1 on September 29, 2017.}
\label{table1}
\setlength{\tabcolsep}{8pt}
\scriptsize
\centering
\begin{tabular}{c c c c c c}
\hline
Time (MJD) & dur. (min) &  $q$ (\%) & $u$ (\%) & $p^*$ (\%) & $J$ intensity light curve\\
\hline
58025.85251 & 22.3 & -0.109 $\pm$ 0.039 & -0.046 $\pm$ 0.044 & 0.10 $\pm$ 0.06 & 0.9956 $\pm$ 0.0016 \\
58025.86697 & 17.8 & 0.020 $\pm$ 0.037 & 0.029 $\pm$ 0.028 & 0.00 $\pm$ 0.05 & 1.0009 $\pm$ 0.0009 \\
58025.87926 & 15.3 & 0.056 $\pm$ 0.032 & -0.006 $\pm$ 0.039 & 0.03 $\pm$ 0.05 & 0.9997 $\pm$ 0.0009\\
58025.88919 & 11.9 & -0.006 $\pm$ 0.039 & 0.085 $\pm$ 0.036 & 0.07 $\pm$ 0.05 & 1.0005 $\pm$ 0.0012\\
58025.89796 & 11.8 & -0.004 $\pm$ 0.044 & -0.003 $\pm$ 0.015 & 0.00 $\pm$ 0.05 & 0.9994 $\pm$ 0.0016\\
58025.90672 & 11.9 & 0.020 $\pm$ 0.046 & 0.040 $\pm$ 0.048 & 0.00 $\pm$ 0.07 & 1.0008 $\pm$ 0.0009\\
58025.91546 & 11.8 & 0.027 $\pm$ 0.045 & 0.049 $\pm$ 0.039 & 0.00 $\pm$ 0.06 & 0.9995 $\pm$ 0.0014\\
58025.92498 & 14.1 & -0.002 $\pm$ 0.032 & 0.033 $\pm$ 0.030 & 0.00 $\pm$ 0.04 & 0.9987 $\pm$ 0.0010\\
58025.93477 & 12.5 & -0.004 $\pm$ 0.015 & -0.013 $\pm$ 0.044 & 0.00 $\pm$ 0.05 & 1.0012 $\pm$ 0.0010\\
58025.94381 & 11.9 & 0.050 $\pm$ 0.022 & 0.019 $\pm$ 0.028 & 0.04 $\pm$ 0.04 & 0.9997 $\pm$ 0.0011\\
58025.95272 & 12.0 & 0.008 $\pm$ 0.056 & -0.031 $\pm$ 0.047 & 0.00 $\pm$ 0.07 & 1.0000 $\pm$ 0.0009\\
58025.96141 & 11.5 & -0.047 $\pm$ 0.049 & -0.082 $\pm$ 0.045 & 0.07 $\pm$ 0.07 & 0.9988 $\pm$ 0.0009\\
58025.96920 & 9.5 & -0.083 $\pm$ 0.043 & -0.079 $\pm$ 0.025 & 0.10 $\pm$ 0.05 & 1.0002 $\pm$ 0.0010\\
58025.97626 & 9.2 & -0.027 $\pm$ 0.042 & 0.059 $\pm$ 0.020 & 0.05 $\pm$ 0.05 & 0.9992 $\pm$ 0.0009\\
58025.98347 & 10.1 & -0.035 $\pm$ 0.083 & 0.055 $\pm$ 0.034 & 0.00 $\pm$ 0.09 & 0.9961 $\pm$ 0.0008\\
58025.99084 & 9.7 & 0.020 $\pm$ 0.024 & 0.082 $\pm$ 0.042 & 0.07 $\pm$ 0.05 & 0.9954 $\pm$ 0.0009\\
58025.99841 & 10.5 & 0.032 $\pm$ 0.026 & 0.011 $\pm$ 0.035 & 0.00 $\pm$ 0.04 & 0.9962 $\pm$ 0.0008\\
58026.00611 & 10.2 & 0.109 $\pm$ 0.021 & 0.052 $\pm$ 0.044 & 0.11 $\pm$ 0.05 & 0.9976 $\pm$ 0.0009\\
58026.01382 & 10.4 & 0.088 $\pm$ 0.029 & 0.028 $\pm$ 0.029 & 0.08 $\pm$ 0.04 & 0.9991 $\pm$ 0.0008\\
58026.02159 & 10.5 & -0.026 $\pm$ 0.047 & 0.025 $\pm$ 0.047 & 0.00 $\pm$ 0.07 & 0.9995 $\pm$ 0.0011\\
58026.02952 & 10.5 & -0.030 $\pm$ 0.041 & 0.006 $\pm$ 0.030 & 0.00 $\pm$ 0.05 & 0.9980 $\pm$ 0.0008\\
58026.03804 & 12.4 & -0.026 $\pm$ 0.038 & 0.073 $\pm$ 0.041 & 0.05 $\pm$ 0.06 & 1.0002 $\pm$ 0.0008\\
58026.05086 & 23.1 & -0.020 $\pm$ 0.024 & 0.043 $\pm$ 0.047 & 0.00 $\pm$ 0.05 & 1.0003 $\pm$ 0.0009\\
58026.06875 & 10.4 & 0.024 $\pm$ 0.038 & 0.015 $\pm$ 0.046 & 0.00 $\pm$ 0.06 & 0.9993 $\pm$ 0.0011\\
58026.07646 & 10.4 & -0.020 $\pm$ 0.046 & 0.051 $\pm$ 0.030 & 0.00 $\pm$ 0.06 & 0.9997 $\pm$ 0.0011\\
58026.08489 & 12.1 & 0.003 $\pm$ 0.033 & 0.089 $\pm$ 0.043 & 0.07 $\pm$ 0.05 & 1.0003 $\pm$ 0.0011\\
58026.09389 & 12.0 & 0.057 $\pm$ 0.036 & 0.014 $\pm$ 0.014 & 0.04 $\pm$ 0.04 & 0.9973 $\pm$ 0.0010\\
58026.10266 & 11.7 & 0.055 $\pm$ 0.040 & 0.023 $\pm$ 0.036 & 0.03 $\pm$ 0.05 & 0.9956 $\pm$ 0.0011\\
58026.11226 & 14.2 & -0.036 $\pm$ 0.062 & -0.030 $\pm$ 0.034 & 0.00 $\pm$ 0.07 & 0.9944 $\pm$ 0.0010\\
58026.12198 & 12.3 & 0.054 $\pm$ 0.044 & -0.016 $\pm$ 0.065 & 0.00 $\pm$ 0.08 & 1.0002 $\pm$ 0.0008\\
\hline
\multicolumn{6}{c}{Best fits for the transits}\\
\hline
 & Trappist-1b & \multicolumn{2}{c}{Trappist-1d} & \multicolumn{2}{c}{Trappist-1e}\\
Mid-time (BJD$_{\rm TDB}$) & 2458026.59164$\pm$2.0$\times$10$^{-3}$ & \multicolumn{2}{c}{2458026.50001$\pm$1.1$\times$10$^{-3}$} & \multicolumn{2}{c}{2458026.35315$\pm$5.6$\times$10$^{-3}$}\\
Transit duration (min) & 40.3$\pm$5.3 & \multicolumn{2}{c}{50.4$\pm$3.0} & \multicolumn{2}{c}{--}\\
Transit depth (\%) & 0.811$\pm$0.080 & \multicolumn{2}{c}{0.444$\pm$0.080} & \multicolumn{2}{c}{0.56$\pm$0.10}\\
\hline
\multicolumn{6}{c}{3 $\sigma$ confidence intervals for the Stokes parameters of Trappist-1}\\
\hline
  & $q$ (\%) & $u$ (\%) & \multicolumn{3}{c}{ }\\
No transit & [-0.026, 0.007] & [-0.018, 0.033] & \multicolumn{3}{c}{ }\\
Transit of Trappist-1b & [-0.040, 0.106] & [ -0.042, 0.037] & \multicolumn{3}{c}{ }\\
Transit of Trappist-1d & [-0.037, 0.099] & [0.018, 0.077] & \multicolumn{3}{c}{ }\\
Transit of Trappist-1e & { [ -0.251, 0.162]} & { [-0.128, 0.112]} & \multicolumn{3}{c}{ }\\
\hline
\end{tabular}
%\begin{minipage}{175.5mm}
%\centering
%\end{minipage}
\end{table*}

\section{Stokes $q$, $u$, $p^*$, and systematics}\label{ap0}

\begin{figure*}
\centering
\includegraphics[width=0.95\textwidth]{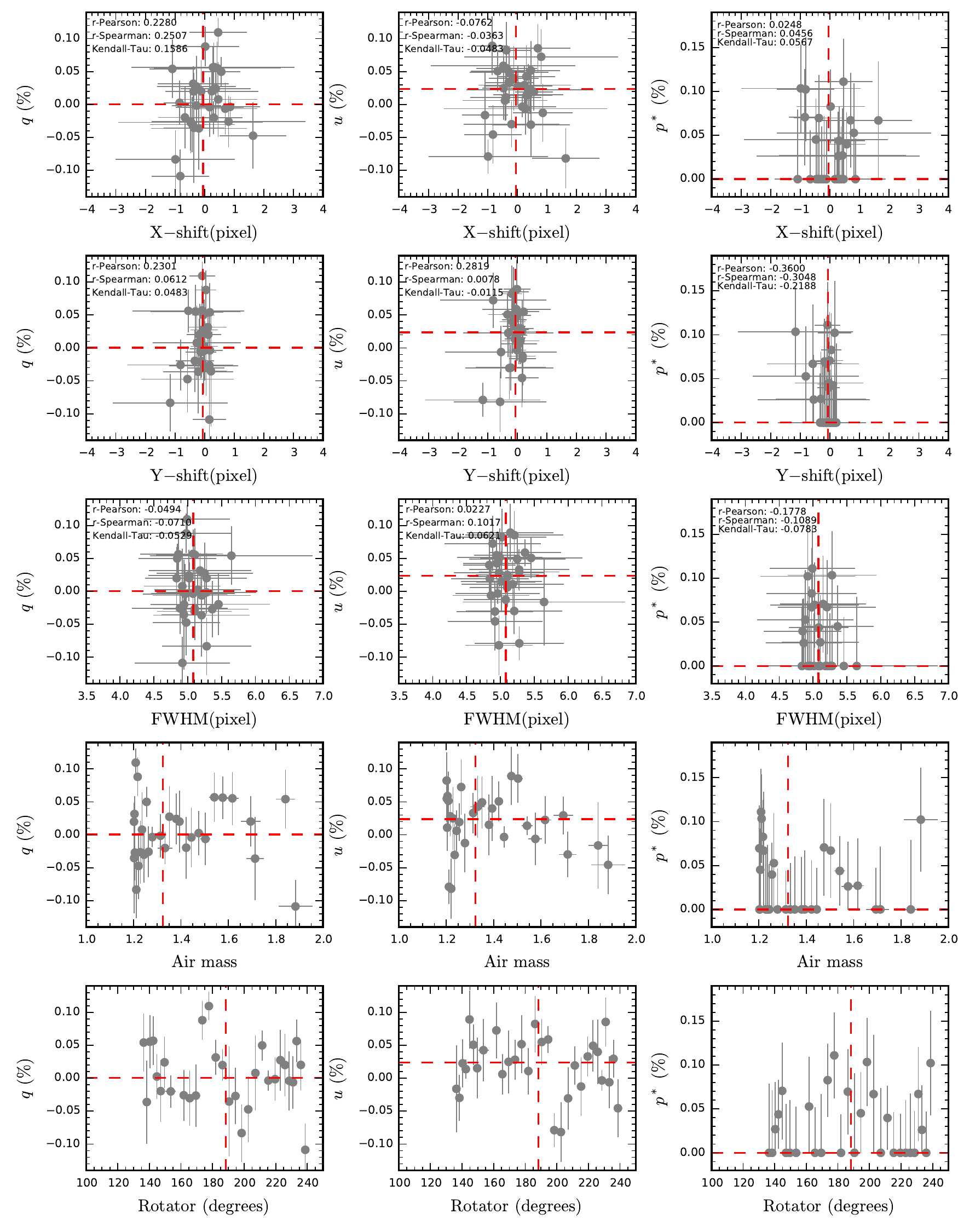}
\caption{From top to bottom, distribution of the Stokes parameters and the degree of linear polarization shown in Figure \ref{fig1} with the centroid of Trappist-1 on the detector,  FWHM, air mass, and telescope rotator angle. Red, dashed lines indicate median values. The values for the $r$ coefficients of correlation by Pearson, Spearman, and Kendall's $\tau$ are also indicated for those cases in which the object's centroid and the FWHM are involved.}
\label{fig_sys}
\end{figure*}

{ Here, we explored the existence of possible systematic effects in the LIRIS linear polarimetric measurements of the night of 2017 September 29 (imaging). In Figure \ref{fig_sys} we plotted the observed $q$, $u$, and $p^*$ as a function of the pixel coordinates of Trappist-1 on the detector, FWHM, air mass, and telescope rotator angle. In all panels, the median values are shown with red dashed lines for comparative purposes. 

The first nine panels of Figure~\ref{fig_sys} display the distribution of our data with the X- and Y-pixel displacement with respect to the median position on the detector and the FWHM (also measured in pixels). To the naked eye, there is no obvious correlation between high and low polarization with any of these parameters. Nevertheless, to further support this statement we computed the Pearson's $r_{\rm p}$, Spearman's $r_{\rm s}$, and the Kendall's $\tau$ coefficients of correlation. The number of observations in our sample ($N$) is large enough to assume that the test statistic ($z$) follows a normal distribution, being computed as $z=\frac{1}{2}\log{\frac{1+r_{\rm p}}{1-r_{\rm p}}}$, $z=r_{\rm s}\sqrt{N-1}$, and $z=\frac{3\tau\sqrt{N(N-1)}}{4N+10}$ for the Pearson's, Spearman's, and Kendall's parameters, respectively. The obtained correlation coefficients are indicated in each of the first nine panels. These values are small, thus suggesting that there is no significant statistical correlation of the polarimetric data with the object's position on the detector and the FWHM (coefficients close to $\pm$1.0 indicate strong correlation). We also performed a classical Fisher's test to investigate the null hypothesis of no correlation \citep{2012msma.book.....F}. We found that the null hypothesis cannot be rejected (i.e., $z>Z_{\alpha/2}$) using typical confidence levels $\ge$95$\%$ ($\alpha\le0.05$) and that the null hypothesis can be discarded with confidence levels of $<90\%$.

As an additional test, we randomly permuted the order of the data and re-computed their $r_{\rm p}$, $r_{\rm s}$, and $\tau$ for a total of 10$^4$ times. If there were any correlation between a particular set of variables, this correlation would disappear when shuffling the data. The distributions of $r_{\rm p}$, $r_{\rm s}$, and $\tau$ for the shuffled data are normally centered around zero, and they are compatible within 1\,$\sigma$ with the quoted $r_{\rm p}$, $r_{\rm s}$, and $\tau$ of the original data. Thus, we discarded spurious polarization being induced by a dependency of $q$, $u$, and $p^*$ on the object's centroid and the determined FWHM.

The bottom six panels of Figure~\ref{fig_sys} display the distribution of our polarimetric data with air mass and the rotator angle of the telescope. Because we are using monitoring data collected in one night, these two parameters reflect the time of the observations. Therefore, the air massses and telescope rotator angles during the planetary transits show higher values of polarization. Furthermore, the observing night was selected because one of the transits (planet d) occurred near culmination, which corresponds to low values of air mass ($\sim$1.2) and telescope rotator angle close to 180 deg. There is no imaging polarimetry of Trappist-1 with telescope rotator angles near 180 deg at moments of no planetary transits on the same observing night. The spectroscopic observations reported in Section~\ref{spectropol} are then useful to assess any systematics: the spectra did not register any planetary transit and show null linear polarization. It is unlikely that linear polarization is correlated with air mass and telescope rotator angle because the imaging data reveal zero polarization immediately before and after the transit of Trappist-1d --- all of these data having similar air masses within 0.1, and because the polarimetric spectra were taken near culmination with similar air mass as the images and with telescope rotator angles between 155$^\circ$ and 171$^\circ$.

}

%If you want to present additional material which would interrupt the flow of the main paper,
%it can be placed in an Appendix which appears after the list of references.

%%%%%%%%%%%%%%%%%%%%%%%%%%%%%%%%%%%%%%%%%%%%%%%%%%

\end{document}